\documentclass[sn-mathphys-num]{sn-jnl}
\usepackage{graphicx}%
\usepackage{multirow}%
\usepackage{amsmath,amssymb,amsfonts}%
\usepackage{amsthm}%
\usepackage{mathrsfs}%
\usepackage[title]{appendix}%
\usepackage{xcolor}%
\usepackage{textcomp}%
\usepackage{manyfoot}%
\usepackage{booktabs}%
\usepackage{algorithm}%
\usepackage{algorithmicx}%
\usepackage{algpseudocode}%
\usepackage{listings}%
\usepackage{draftwatermark}

\SetWatermarkText{Uncorrected Proof}

\SetWatermarkScale{1.5}
\usepackage{tikz}
\usepackage{physics}
\usepackage{siunitx}
\usepackage{hyperref}
\hypersetup{colorlinks=true,
citecolor=black,
filecolor=black,
linkcolor=black,
urlcolor=black}
\usepackage[version=4]{mhchem}
\usepackage{subcaption}

\DeclareSIUnit\pcm{pcm}
\DeclareSIUnit\year{an}
\DeclareSIUnit\day{j}
\DeclareSIUnit\barn{b}
\DeclareSIUnit\atm{atm}

\newcommand{\keff}{k_{\text{eff}}}
\usepackage{acro}
\DeclareAcronym{DNP}{
	short = DNPs ,
	long = Delayed Neutron Precursors
}
\DeclareAcronym{UF6}{
	short = UF$_{6}$ ,
	long = uranium hexafluoride
}

\begin{document}

\title[One-dimensional gas-fueled nuclear reactor with thermal feedback]{One-dimensional gas-fueled nuclear reactor with thermal feedback}

\author*[1]{Mathis Caprais}\email{mathis.caprais@universite-paris-saclay.fr}

\author[1]{Kacim François-Elie}\email{kacim.francois-elie@universite-paris-saclay.fr}

\author[2]{Daniele Tomatis}

\affil*[1]{\orgname{Université Paris-Saclay}, \orgaddress{\city{Gif-sur-Yvette}, \postcode{91190}, \country{France}}}


\affil[2]{\orgaddress{\street{Via Almese 15}, \city{Torino}, \postcode{10129}, \country{Italy}}}


\abstract{This study explores a simplified one-dimensional subchannel of a graphite-moderated nuclear reactor operating with a gaseous core in steady-state conditions, reproducing a neutronic-thermal-fluid-dynamics coupled problem with thermal feedback. The fuel gas, consisting of a homogeneous mixture of uranium hexfluoride (\ce{UF_6}) and helium, is assumed to be ideal, with simplifications made to its thermodynamic state. Due to the high thermal expansion of the fuel, a possible interesting strong coupling is anticipated. The discrete ordinates' method is used to compute the one group scalar flux, the effective multiplication factor and the power released by the core. Six groups of Delayed Neutron Precursors (DNPs) are used to take into account the fuel motion drift. Compressible Euler equations are solved with a monolithic approach and the two-physics problem is treated with Picard iterations.
	As expected, the effective multiplication factor of a subchannel is shown to increase with the inlet pressure. The critical pressure, representing the threshold at which the system achieves criticality, changes as the fuel mixture changes. High thermal feedback coefficients are observed due to the high thermal expansion of the fuel. The amount of helium in the mixture greatly affects the temperature at core outlet in critical configurations. This study shows that a gaseous fuel reactor can be brought to criticality varying the inlet pressure. The thermal feedback is strong and should be taken into account in the design of the system.}

\maketitle

\section*{Introduction}
Gas and Vapor Core Reactors (G/VCR) are nuclear fission reactors using a mixture of fissile gas as a fuel. These reactors were often considered using a uranium fluoride gas \ce{UF_n}, $n=1$ to $6$, mixed with other metallic fluoride and helium. These systems were extensively studied between the 1950s and the 1980s \cite{klein1987thermodynamic, thom1977gaseous}.

A gaseous fuel offers the possibility of very high working temperatures \cite{kuijper1992calculational, clement1976analysis} (from thousands to ten of thousands of degrees \cite{van1983physics, thom1977gaseous}) and also more direct ways to convert the energy released by fission into electricity such as magneto-inductive or magneto-hydrodynamic conversion \cite{kuijper1992calculational, shubov2021gas, klein1987thermodynamic, van1983physics, clement1976analysis, gffr_power_cycle}. Gas core reactors also offer a homogeneous burn-up, continuous reprocessing of the fuel and a lower mass of fissile materials than traditional reactors \cite{kuijper1992calculational, thom1977gaseous}. Some of these advantages were observed in a demonstrator built and operated in the Soviet Union, using \ce{UF_6} as a fuel \cite{clement1976analysis}. The fuel was enriched at $90\%$ in \ce{^{235}U}, flowing through beryllium-moderating channels surrounded by a graphite reflector \cite{clement1976analysis}. Refueling was done continuously. The reactor demonstrated large negative feedback coefficients due to the high thermal expansion of the fissile gas. 
The studies previously cited were interested in the performance of the reactor rather than the coupling between neutronic and thermodynamic of a compressible and flowing neutron multiplying gas. Especially, this research paper investigates the influence of the fissile gas pressure on its criticality, in a coupled physics framework.

In this study the coupling between neutronics and thermodynamics of a neutron multiplying gas is investigated as the classical problem of the reactor subchannel \cite{tomatis2019reformulation}. In Sec. \ref{sec:thermo_equations}, the main equations describing the gaseous fuel are presented. The reactor subchannel is studied in the steady-state. In Sec. \ref{sec:neutron_balance}, neutron balance equation are presented taking into account the drift of Delayed Neutron Precursors (DNPs) due to circulation of the gas. All equations are discretized using finite volumes on an equidistant mesh. A one energy group deterministic solver of the neutron transport equation has been developed for fast calculations of the coupled problem in a simplified geometry. This allows for a fast exploration of the parameters space, and avoid coupling a costly Monte-Carlo code to a CFD solver. Compressible Euler's equations are solved monolithically using Newton's algorithm, with the Jacobian matrix computed analytically. The thermal and pressure feedback on cross-sections is taken into account using the state law of the fuel. The coupled system is solved using Picard's iterations. The macroscopic cross-sections of the nuclear fuel are generated using the OpenMC Monte-Carlo code \cite{romano2015openmc}. The evolution of the effective multiplication factor as a function of pressure, temperature distributions and feedback coefficients are presented in Sec. \ref{sec:results} for different fuel mixtures.
\section{Physical model}
\subsection{Thermodynamic equations of gaseous fuel}\label{sec:thermo_equations}
In this section, the balance equations and closure law for a gas mixture are given.
\subsubsection{Physical parameters of interest}
The fuel gas is assumed to be ideal and composed of a mixture of helium and \ce{UF_6}, with mixing atomic fraction $e$. \ce{UF_6} is chosen as the fissile gas as it had already been used in criticality experiments. Ideality greatly simplifies the equations describing the temperature ($T$, \SI{}{\kelvin}), velocity ($u$, \SI{}{\meter\per\second}), density ($\rho$, \SI{}{\kilo\gram\per\cubic\meter}) and pressure of the fuel ($p$, \SI{}{\pascal}). The ideal gas assumption also provides a closure law for compressible Euler's equations \cite{chorin1990mathematical}. The ideal gas law states that,
\begin{equation}
	p = \rho R_s T,
\end{equation}
$R_s = R / M$ is the gas specific constant (\SI{}{\joule\per\kelvin\per\kilo\gram}) with $R = \SI{8.314}{\joule\per\kelvin\per\mole}$ the ideal gas constant and $M$ the gas molar mass (\SI{}{\kilo\gram\per\mole}). The ideal gas law and all the following properties of the mixture can be derived from statistical physics assuming the energy of the molecules is purely kinetic \cite{Diu_Guthmann_Lederer_Roulet_2012}. Even though \ce{UF_6} is a complex molecule with internal degrees of freedom, the fuel is assumed ideal at high pressure and temperature as the literature is scarce on the evaluation of the thermophysical properties of the gas in this range of pressure and temperature. The ideal gas law preserves the high thermal expansion of the fuel of interest for assessing the thermal feedback. The average molar mass of the mixture is defined as,
\begin{equation}
	M = e M_{\ce{UF_6}} + \qty(1-e) M_{\ce{He}}.
\end{equation}
It is assumed that the heat capacity ratio,
\begin{equation}
	\gamma = \frac{C_p}{C_v},
\end{equation}
is known for both gases. This ratio tells us how heat is used in the gas. At constant volume ($C_v$) all the energy is converted into a temperature increase while at constant pressure ($C_p$) the system shall be free to expand. This leaves less energy to raise the temperature of the gas and thus $C_p > C_v$ so $\gamma > 1$. Therefore, a gas with a heat capacity ratio larger than unity will expand more when receiving the same amount of energy than a gas with $\gamma \simeq 1$. The heat capacity ratio of the mixture is calculated using Mayer's relation for the mixture,
\begin{equation}
	C_p - C_v = R,
	\label{eq:mayer_relation}
\end{equation}
the average heat capacity ratio $\bar{\gamma}$ as,
\begin{equation}
	\bar{\gamma} = 1 + \frac{1}{\frac{e}{\gamma_{\ce{UF_6}} - 1} + \frac{(1-e)}{\gamma_{\ce{He}} - 1}} \qq{and} c_v = \frac{R_s}{\bar{\gamma} - 1}.
	\label{eq:cv_mixture}
\end{equation}
Higher helium content reduces the constant volume heat capacity per unit of volume $c_v = C_v / \rho$, thus allowing higher temperature at the outlet to preserve the rated thermal power of the system. The quantities used in the equations above take the values listed in Table \ref{table:thermo_data} in the following.
\begin{table}[ht]
	\centering
	\caption{Physical constants of helium and uranium hexafluoride.}
	\vspace{5mm}
	\begin{tabular}{ccc}
		\hline
		Parameter            & Value                              & Reference                               \\
		\hline
		\hline
		$M_{\ce{He}}$        & $\SI{4e-3}{\kilo\gram\per\mole}$   & -                                       \\
		$M_{\ce{UF_6}}$      & $\SI{352e-3}{\kilo\gram\per\mole}$ & -                                       \\
		$\gamma_{\ce{He}}$   & $5/3$                              & \cite{Diu_Guthmann_Lederer_Roulet_2012} \\
		$\gamma_{\ce{UF_6}}$ & $1.062$                            & \cite{dewitt1960uranium}                \\
		\hline
	\end{tabular}
	\label{table:thermo_data}
\end{table}
The heat capacity ratio of \ce{UF_6} is a function of temperature. Over the range of high pressures and temperatures of this study, its heat capacity ration varies from $1.08$ to $1.062$ \cite{uf6_heat_cap}, which represents a relative variation of less than \SI{2}{\percent}. Therefore, heat capacity ratio of \ce{UF_6} is assumed to be constant.
\subsubsection{Balance Equations}
The evolution of the mass, momentum and energy of the gas is described by the steady-state Euler's equations, written as a continuity equation,
\begin{equation}
	\div{\vb{F}\qty(\vb{X})} = \vb{S} \qq{with}\vb{S} =
	\begin{pmatrix}
		0   \\
		0   \\
		P_v \\
	\end{pmatrix}
	\qq{and}
	\vb{X} = \begin{pmatrix}
		\rho \\
		j    \\
		\varepsilon
	\end{pmatrix}
	\label{eq:coupled_steady_euler}
\end{equation}
where $\vb{S}$ is a source of mass, momentum and energy. $P_v$ is the fission volume heat source. $\vb{X}$ is the conservative vector containing the density $\rho$, the volume momentum $j=\rho u$ and the volume total energy $\varepsilon$. The operator $\vb{F}$ acting on the vector $\vb{X}$ is \cite{chorin1990mathematical},
\begin{equation}
	\vb{F}\qty(\vb{X}) =
	\begin{pmatrix}
		j                    \\
		\frac{j^2}{\rho} + p \\
		\qty(\varepsilon + p) \frac{j}{\rho}
	\end{pmatrix}
	=
	\begin{pmatrix}
		j                                                                                        \\
		\frac{j^2}{\rho} + \qty(\bar{\gamma} - 1)\qty(\varepsilon - \frac{1}{2}\frac{j^2}{\rho}) \\
		\qty(\varepsilon + \qty(\bar{\gamma} - 1)\qty(\varepsilon - \frac{1}{2}\frac{j^2}{\rho})) \frac{j}{\rho}
	\end{pmatrix}
	\label{eq:flux_vector}
\end{equation}
where the pressure has been eliminated using the ideal gas law relating conservative quantities. The non-conservative variables (temperature, pressure, velocity) are deduced from the conservative variables using the following relations,
\begin{equation}
	u = \frac{j}{\rho}, \quad p = \qty(\bar{\gamma} - 1)\qty(\varepsilon - \frac{1}{2}\frac{j^2}{\rho}), \quad T = \frac{1}{\rho c_v}\qty(\varepsilon - \frac{1}{2}\frac{j^2}{\rho}).
	\label{eq:non_conservative_variables}
\end{equation}
\subsection{Neutron balance equation}\label{sec:neutron_balance}
In a gaseous fuel the diffusion approximation is not valid as scattering macroscopic cross-sections are of the same order of magnitude as absorption macroscopic cross-sections. Therefore, the one-dimensional one-group transport equation is solved using the discrete ordinates or $S_N$ method. The angular dimension is treated using the Gauss-Legendre quadrature, that is with the cosines of the angles given as roots of the Legendre polynomials, and weights selected to preserve their integrals over $\left[-1, 1\right]$ \cite{abramowitz1968handbook}. The sum of $\omega_i$ is taken to be equal to $2$ as $4\pi = \int_{4\pi}\dd{\vb*{\Omega}} = 2\pi \int_{-1}^{1}\dd{(\cos \theta)}\simeq 2\pi \sum_i \omega_i$. The scalar flux $\phi$ is calculated as the angle integral of the angular flux $\psi$,
\begin{equation}
	\phi (x) = 2\pi \int_{-1}^{1}\dd{\mu}\psi(x, \mu) \simeq 2\pi \sum_i \omega_i \psi(x, \mu_i).
\end{equation}
The balance equation for the angular flux $\psi$ for the $i$-th direction sampled is,
\begin{equation}
	\mu_i \pdv{\psi_i}{x} + \Sigma_t \psi_i = \frac{\Sigma_s}{2}\sum_k \omega_k \psi_k + q_{\mathrm{ext}}
\end{equation}
where $\Sigma_t$ is the total macroscopic cross-section, $\Sigma_s$ is the scattering macroscopic cross-section (anisotropic scattering is neglected). The source $q_{\mathrm{ext}}$ is given as:
\begin{equation}
	q_{\mathrm{ext}} = \qty(1-\beta)\frac{\nu}{\keff}\frac{\Sigma_f}{2}\sum_k \omega_k \psi_k + \sum_j \frac{\lambda_j}{4 \pi} C_j,
\end{equation}
where $\Sigma_f$ and $\nu$ are respectively the macroscopic fission cross-section and the neutron multiplicity. The effective multiplication factor $\keff$ is playing the role of eigenvalue for the homogeneous problem. The total delayed neutron fraction $\beta = \sum_j \beta_j$ is the sum of the delayed neutron fractions. The precursor concentration $C_j$ in the $j$ DNPs group is given as a solution of the transport equation:
\begin{equation}
	\pdv{uC_j}{x} + \lambda_j C_j = \beta_j \frac{\nu}{\keff}\frac{\Sigma_f}{2} \sum_k \omega_k \psi_k,
\end{equation}
where $\beta_j$ and $\lambda_j$ are respectively the DNPs proportion and the decay constant of the $j$-th precursors group. The concentration of DNPs at the inlet is set to be zero meaning that fresh fuel constantly enters the reactor, or that the recirculation time is very long. Vaccuum boundary conditions are imposed at the edges of the slab for the angular flux
\begin{equation}
	\psi \qty(0, \mu > 0) = \psi \qty(\ell, \mu < 0) = 0.
\end{equation}
\section{A case study}
\subsection{Reference Configuration}\label{sec:reference_configuration}
\begin{figure}[h]
	\centering
	\begin{subfigure}[b]{0.45\textwidth}
		\includegraphics[width=\textwidth]{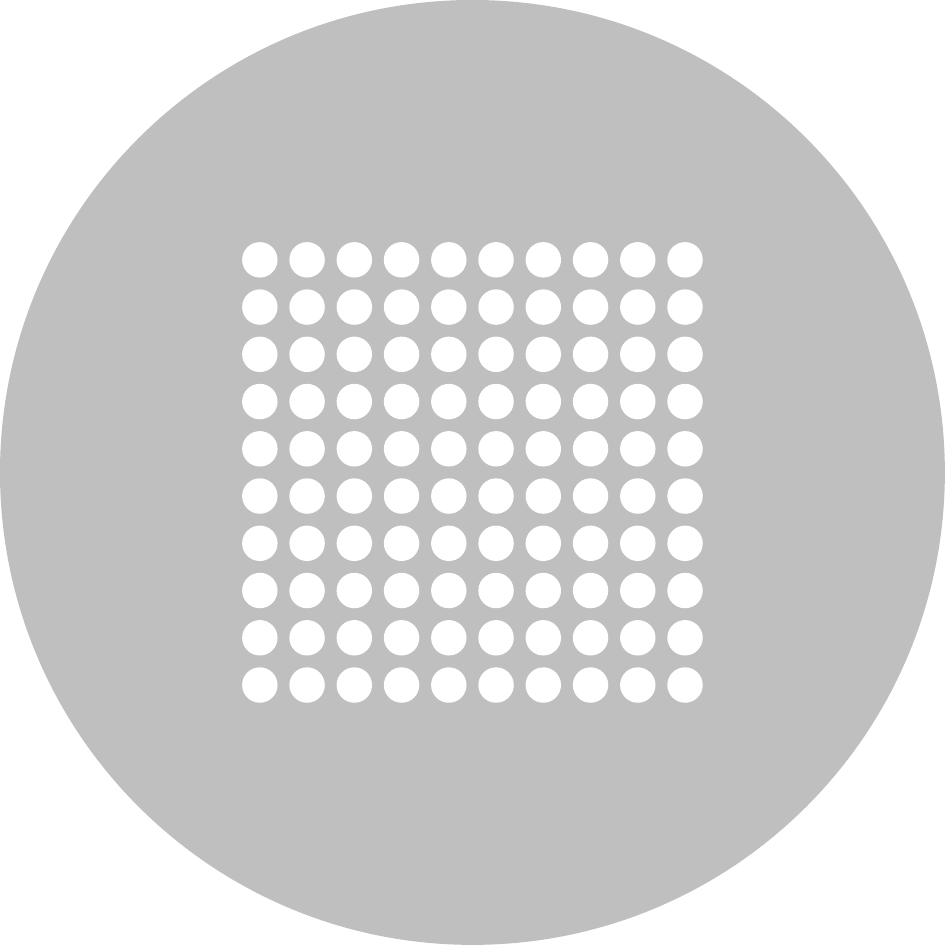}
		\caption{Full Reactor}
		\label{fig:full_reactor}
	\end{subfigure}
	\hfill
	\begin{subfigure}[b]{0.45\textwidth}
		\includegraphics[width=\textwidth]{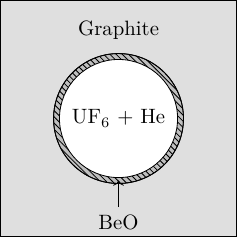}
		\caption{Subchannel}
		\label{fig:subchannel}
	\end{subfigure}
	\caption{The complete reactor and one of the reactor subchannels.}
	\label{fig:figures}
\end{figure}

The reactor studied is shown in Fig. \ref{fig:full_reactor}. It is a \SI{5}{\meter} high cylinder of nuclear graphite with density \SI{1.82e3}{\kilo\gram\per\cubic\meter} composed of $N_{\mathrm{channels}}=\num{100}$, \SI{20}{\centi\meter} by \SI{20}{\centi\meter} subchannels shown in Fig. \ref{fig:subchannel}. Each subchannel is a tube of \SI{10}{\centi\meter} in diameter surrounded by a \SI{0.5}{\centi\meter} cylindrical shell of \ce{BeO}. Beryllium oxide serves both as refractory material \cite{higgins2010materials} and a neutron reflector between the graphite channel and the gaseous fuel. The height was chosen to ensure criticality. The subchannel diameter was chosen in order to balance moderation and increased velocity through thermal feedback. The inlet velocity is set to be \SI{20}{\meter\per\second}, the inlet pressure \SI{40}{\atm}\footnote{$\SI{1}{\atm}=\SI{101325}{\pascal}$} and the inlet temperature \SI{600}{\kelvin}. The reference thermal power of the system is $P_{\mathrm{th}}=\SI{3}{\giga\watt}$.
\subsection{Fuel}
One-group macroscopic cross-sections are prepared by Monte Carlo calculations that reproduces the working conditions of a subchannel in the reactor.
The subchannel is filled with a mixture of $50\%$\ce{UF_6}-$50\%$\ce{He}. The gas is enriched at $3\%$ in \ce{^{235}U}. Reflective boundary conditions are imposed on the boundaries of the subchannel. The simulation is conducted using OpenMC 0.13.3 at nominal parameters \cite{romano2015openmc}. OpenMC is an open source Monte-Carlo transport code developed by researchers from the Computational Reactor Physics Group at the MIT. It is capable of solving multiple kind of neutron and photon transport problems. It is convenient to use OpenMC to generate multigroup cross-sections libraries, which can then be used in deterministic codes such as OpenMOC \cite{BOYD201443}. The nuclear data library used is ENDF-VIII0 \cite{brown2018endf}. $50$ batches are simulated, the first $10$ being inactive. $10000$ particles are simulated per batch. Six DNPs groups are used. Cross-sections are then calculated at any temperature and pressure using the ideal gas law, such as:
\begin{equation}
	\Sigma \qty(T,p) = \Sigma \qty(T_0,p_0) \frac{p T_0}{p_0 T}.
	\label{eq:sigma_p_t}
\end{equation}
Macroscopic cross-sections are expected to increase with pressure and decrease with temperature. Eq. \eqref{eq:sigma_p_t} assumes that the microscopic cross-sections do not change significantly with different thermodynamic states. The validity of the assumptions leading to Eq. \eqref{eq:sigma_p_t} is tested by varying the mixture's temperature, showing an agreement within \SI{1}{\percent} with the Monte Carlo results. Density variations showed larger differences instead, up to \SI{40}{\percent}. Results for a validation covering a larger range of thermodynamic states will be carried as future development.
\subsection{Full-core criticality calculation}
The reactor depicted in Fig. \ref{fig:full_reactor} reaches criticality, $\keff = \SI{1.0110\pm 0.0004}{}$, at \SI{50}{\atm} with an equimolar fuel. Vaccuum boundary conditions are imposed at the surfaces of the graphite cylinder. The effective multiplication factor was found to be an increasing function of the fuel pressure. Increasing the pressure increases the mass of fissile materials within the subchannels ultimately increasing the overall reactivity.
\section{Discretization of the equations}
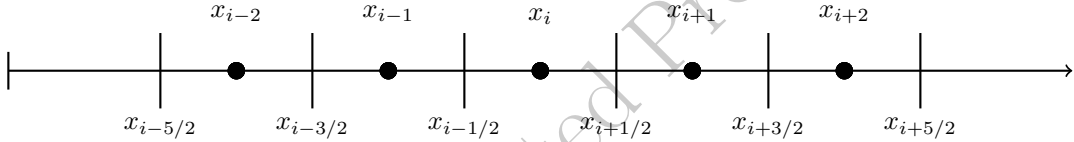
\begin{figure}[h]
	\centering
	\begin{tikzpicture}
		\draw[thick, ->] (-2,0) -- (12,0);
		\draw[thick] (-2,-0.25) -- (-2,0.25);
		\foreach \x in {0,2,4,6,8,10}{
				\draw[thick] (\x,0.5) -- (\x,-0.5);
			}
		\foreach \x in {1,3,5,7,9}
		\foreach \x in {1,3,5,7,9}{
				\draw[thick, fill=black] (\x,0) circle (0.1);
			}
		\node[below] at (0,-0.5) {$x_{i-5/2}$};
		\node[below] at (2,-0.5) {$x_{i-3/2}$};
		\node[below] at (4,-0.5) {$x_{i-1/2}$};
		\node[below] at (6,-0.5) {$x_{i+1/2}$};
		\node[below] at (8,-0.5) {$x_{i+3/2}$};
		\node[below] at (10,-0.5) {$x_{i+5/2}$};
		\node[above] at (1,0.5) {$x_{i-2}$};
		\node[above] at (3,0.5) {$x_{i-1}$};
		\node[above] at (5,0.5) {$x_{i}$};
		\node[above] at (7,0.5) {$x_{i+1}$};
		\node[above] at (9,0.5) {$x_{i+2}$};
	\end{tikzpicture}
	\caption{Discretized segment}
	\label{fig:discretized_segment}
\end{figure}

Let $\Delta x>0$, the segment representing the reactor core is subdivided into an equidistant sequence of $\mathbf{N}$ intervals $K_i$ defined by
$$
	K_i=\left(x_{i-1 / 2}, x_{i+1 / 2}\right), \quad x_{i+1 / 2}=i \Delta x, \quad \forall i \in \mathbb{Z},
$$
as shown in Fig. \ref{fig:discretized_segment}. The center $x_i$ of the cell $K_i$ is $x_i=\left(x_{i+1 / 2}+x_{i-1 / 2}\right) / 2, x_{i-1 / 2}$ and $x_{i+1 / 2}$ mark the left and right faces of the cell $i$, respectively.

\subsection{Discretized neutron balance equation}
The neutron balance equation is integrated in every cell, in which the cross-sections are uniform. Spatial integration yields:
\begin{equation}
	\frac{1}{\Delta x}\int_{x_{i-1 / 2}}^{x_{i+1 / 2}}\dd{x}\Sigma \psi = \Sigma_i \psi_i,
\end{equation}
where the reaction and the angular direction indices are dropped for simplicity. The spatial gradient of the angular flux is approximated using the upwind approximation,
\begin{equation}
	\frac{1}{\Delta x}\int_{x_{i-1 / 2}}^{x_{i+1 / 2}}\dd{x}\mu \dv{\psi}{x} \simeq \frac{\mu}{\Delta x} (\psi_i - \psi_{i-1}).
\end{equation}
While integrating the DNPs balance equation, the upwind scheme is used for the advection term:
\begin{equation}
	\frac{1}{\Delta x}\int_{x_{i-1 / 2}}^{x_{i+1 / 2}}\dd{x} \dv{}{x}uC \simeq \frac{u_i C_i - u_{i-1}C_{i-1}}{\Delta x}, \qq{and} \frac{1}{\Delta x}\int_{x_{i-1 / 2}}^{x_{i+1 / 2}}\dd{x} \lambda C = \lambda C_i,
\end{equation}
where again the index of precursor family is dropped for simplicity. Cell quantities are considered as volume-averaged. It is assumed that $u_i C_i$ is almost equal to the average DNPs mass flux in the $i$-th cell $(uC)_i$.
\subsection{Discretization of compressible Euler equations}
The flux vector is integrated on the volume of a cell such as,
\begin{equation}
	\frac{1}{\Delta x}\int_{x_{i-1 / 2}}^{x_{i+1 / 2}}\dd{x} \dv{}{x}\vb{F}(\vb{X}) = \frac{\vb{F}_{i+1 / 2} - \vb{F}_{i-1 / 2}}{\Delta x} \simeq \frac{\vb{F}_{i} - \vb{F}_{i-1}}{\Delta x},
\end{equation}
where the upwind scheme has been used.
\section{Numerical Scheme}
The discretized equations presented above, and the coupling schemes presented hereafter are implemented in Python 3.10 with NumPy 1.26.1.
\subsection{Monolithic coupling of Euler's equations}
Euler's equations are solved as a fixed point problem. The source vector is moved on the left-hand side of the balance equation \eqref{eq:coupled_steady_euler}, such as $\vb{P}(\vb{X}) = \div{F\qty(\vb{X})} - \vb{S} = \vb{0}$ and the fixed-point problem is solved using Newton's method. Non-homogeneous Dirichlet boundary conditions are applied for the pressure, temperature and velocity:
\begin{equation}
	p(x=0) = p_0, \quad T(x=0) = T_0, \quad u(x=0) = u_0 \qq{so that} \rho(x=0)=\frac{p_0}{R_s T_0}.
\end{equation}
These boundary conditions are written in conservative variables:
\begin{equation}
	j (x=0) = \frac{u_0 p_0}{R_s T_0} \qq{and} \varepsilon (x=0) = \frac{p_0}{\bar{\gamma} - 1} + \frac{1}{2}\frac{u_0^2 p_0}{R_s T_0}.
\end{equation}
In the following equations, $K_x$ denotes the general first order discrete derivative operator, such as:
\begin{equation}
	K_x = \frac{1}{\Delta x}
	\begin{bmatrix}
		1      & 0      & 0      & 0      & \dots  & 0      \\
		-1     & 1      & 0      & 0      & \dots  & 0      \\
		0      & -1     & 1      & 0      & \dots  & 0      \\
		0      & 0      & -1     & 1      & \dots  & 0      \\
		\vdots & \vdots & \vdots & \vdots & \ddots & \vdots \\
		0      & 0      & 0      & \dots  & -1     & 1      \\
	\end{bmatrix}
\end{equation}
and the bold versions of the conservatives fluxes denote their discretized versions along the $x$-axis. The Jacobian of the system is:
\begin{equation}
	J(\vb{X}) =
	\begin{bmatrix}
		0_{N\times N}                                                                                                                                                 & K_x                                                                                                                                     & 0_{N\times N}                                  \\
		\vphantom{0}                                                                                                                                                  & \vphantom{0}                                                                                                                            & \vphantom{0}                                   \\
		\frac{\vb{j}^2}{\vb*{\rho}^2}\frac{\bar{\gamma} - 3}{2}\odot K_x                                                                                              & \frac{\vb{j}}{\vb*{\rho}}\qty(3-\bar{\gamma})\odot K_x                                                                                  & \qty(\bar{\gamma} - 1) K_x                     \\
		\vphantom{0}                                                                                                                                                  & \vphantom{0}                                                                                                                            & \vphantom{0}                                   \\
		\qty(\bar{\gamma} \frac{\boldsymbol{\varepsilon}}{\vb*{\rho}} - \qty(\bar{\gamma} - 1)\frac{\vb{j}^2}{\vb*{\rho}^2})\qty(-\frac{\vb{j}}{\vb*{\rho}})\odot K_x & \qty(\bar{\gamma} \boldsymbol{\varepsilon} - \frac{3}{2}\qty(\bar{\gamma} - 1)\frac{\vb{j}^2}{\vb*{\rho}})\frac{1}{\vb*{\rho}}\odot K_x & \bar{\gamma}\frac{\vb{j}}{\vb*{\rho}}\odot K_x \\
		\vphantom{0}                                                                                                                                                  & \vphantom{0}                                                                                                                            & \vphantom{0}                                   \\
	\end{bmatrix}
\end{equation}
where $\odot$ is the element-wise or Hadamard product. The updated vector of conservative variables at the $(n+1)$-iteration is given by,
\begin{equation}
	\vb{X}^{(n+1)} = \vb{X}^{(n)} + \delta\vb{X}^{(n)},
\end{equation}
and the correction $\delta\vb{X}^{(n)}$ is given by:
\begin{equation}
	J(\vb{X}^{(n)})\delta\vb{X}^{(n)} = - \vb{P}(\vb{X}^{(n)}).
\end{equation}
Newton's algorithm is considered converged when:
\begin{equation}
	\frac{\norm{\vb{X}^{(n+1)} - \vb{X}^{(n)}}_\infty}{\norm{\vb{X}^{(n)}}_\infty} < \num{1e-6}.
\end{equation}
\subsection{Coupling with neutronics}
\begin{figure}[hbtp]
	\centering
	\includegraphics[width=60mm]{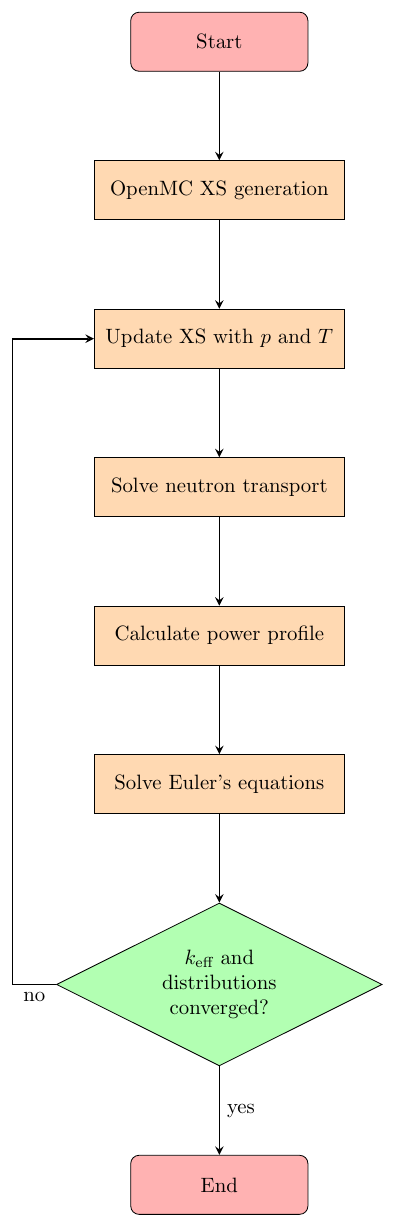}
	\caption{Coupling scheme of the cross-section generation, thermodynamic and neutronic codes.}
	\label{fig:coupling_scheme}
\end{figure}

The coupling scheme to solve the coupled thermodynamic and neutronic problem is presented in Fig. \ref{fig:coupling_scheme}.
When the conservative vector $\vb{X}$ is obtained, the pressure, velocity and temperature are evaluated with Eq. \eqref{eq:non_conservative_variables}. Then, the angular flux is calculated for the $N$ angular directions with the updated macroscopic cross-sections. Each component of the angular flux is computed until convergence on the scattering source is attained. Power iterations are then performed on the fission source which accounts for both prompt and delayed neutrons. The norm of the scalar flux is imposed knowing the thermal power of the channel,
\begin{equation}
	P_{\mathrm{th}} = \frac{1}{N_{\mathrm{channels}}}\int \dd{V} P_v =\frac{1}{N_{\mathrm{channels}}}\int \dd{V} \kappa \Sigma_f \phi,
\end{equation}
where $\kappa \Sigma_f$ (\SI{}{\joule\per\meter}) is the energy released by a fission event times the macroscopic fission cross-section. Power normalization must be ensured at each iteration before solving the Euler equations. Both solvers exchange information until convergence on the effective multiplication factor and the neutronic vector $\vb{Y}$ containing the scalar flux and precursors concentrations is reached,
\begin{equation}
	\norm{1 - \frac{\keff^{(n+1)}}{\keff^{(n)}}} < \num{1e-6} \qq{and} \frac{\norm{\vb{Y}^{(n+1)} - \vb{Y}^{(n)}}_\infty}{\norm{\vb{Y}^{(n)}}_\infty} < \num{1e-6}.
\end{equation}
\section{Results}\label{sec:results}
The following results were obtained with $16$ angular directions and $500$ spatial cells.

\subsection{Reactivity as a function of pressure}
\begin{figure}[ht]
	\centering
	\includegraphics[width=100mm]{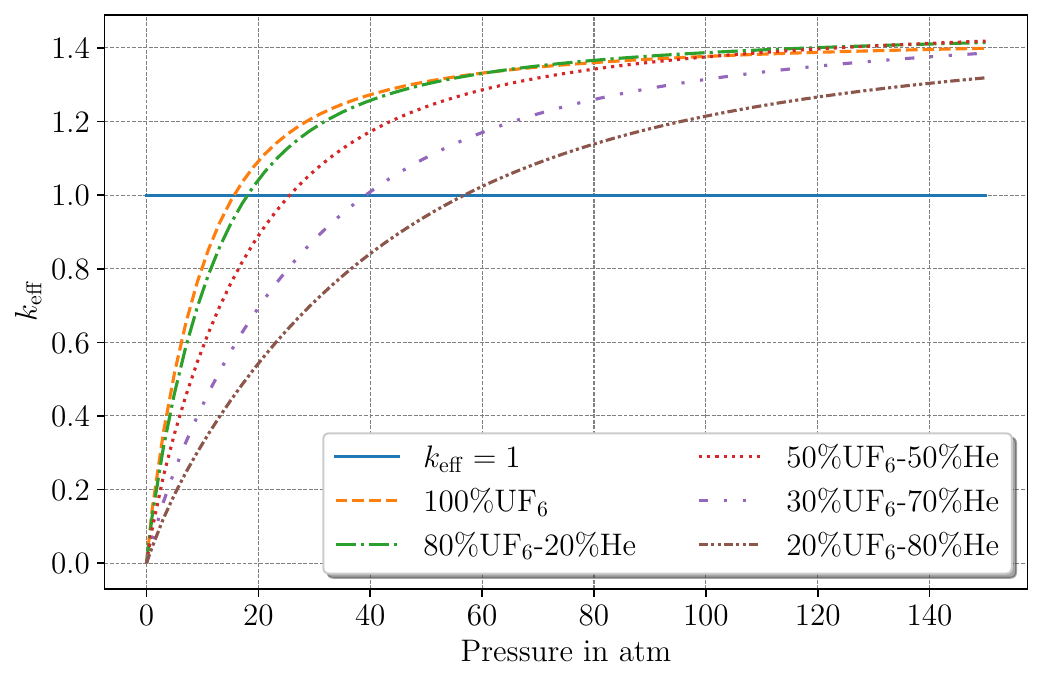}
	\caption{Evolution of the effective multiplication factor as a function of the inlet pressure}
	\label{fig:keff_vs_p}
\end{figure}

Higher pressure while keeping the inlet temperature and the thermal power unchanged shows a flow with higher density gas along the channel, which in turn yields higher multiplication factor due to the cross-section model from Eq. \eqref{eq:sigma_p_t}. This behavior can be noticed in Fig. \ref{fig:keff_vs_p} for different fuel mixtures.
As anticipated, the shift from subcritical to critical state takes place at a reduced inlet pressure when employing a pure fissile gas, in contrast to a gas mixture with a higher proportion of helium. Higher helium content in the mixture results in higher temperature justifying the slower rate in achieving criticality.
Moreover, still as per Eq. \eqref{eq:sigma_p_t} cross-sections decrease when temperature increases, explaining further the trend of the multiplication factor in the plot.
\subsection{Temperature distributions}
\subsubsection{Non-critical temperature distributions}
\begin{figure}[ht]
	\centering
	\includegraphics[width=100mm]{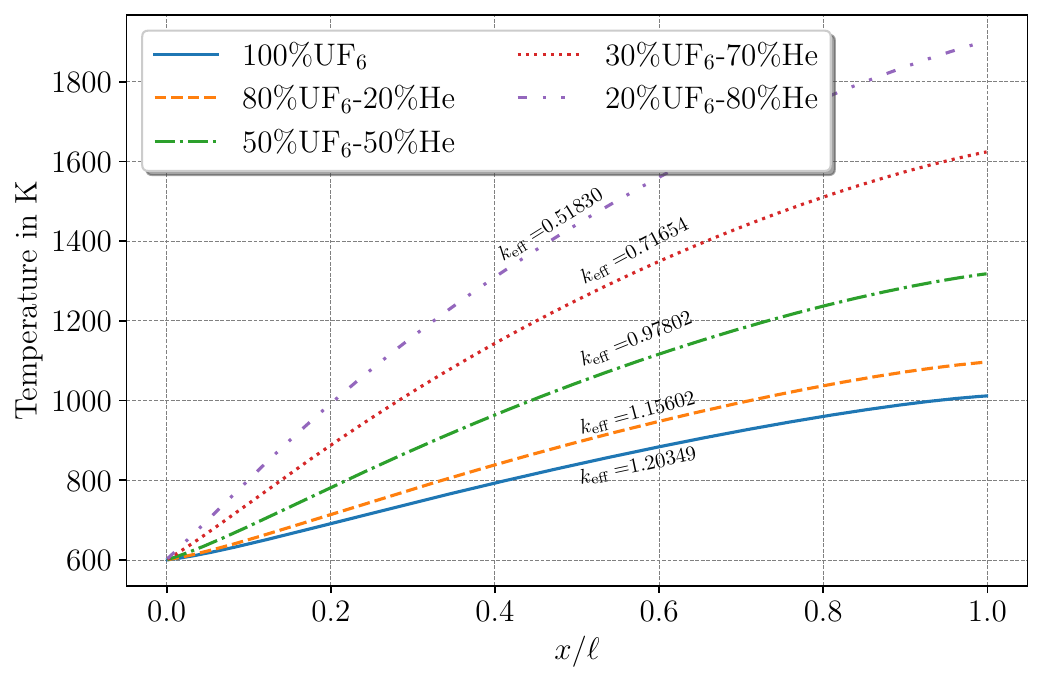}
	\caption{Evolution of the temperature in the subchannel for different fuel composition. The effective multiplication factor is given for each curve.}
	\label{fig:temperature_mixture}
\end{figure}

Fig. \ref{fig:temperature_mixture} shows the distributions of temperatures for different proportions of helium in the fuel mixture.
As pointed out in the derivation of the main equations, an increase in the helium content in the fuel reduces the heat capacity of the mixture. This turns out achieving higher temperature at the core outlet. Fig. \ref{fig:temperature_mixture} shows that very high temperature gradients can be obtained, possibly leading to higher thermal efficiency of the conversion system. However, the amount of helium should be adjusted to obtain criticality at a viable pressure level according to the technological constraints of the system. The systems with higher helium amount are farer from criticality, provided that the inlet conditions do not change. The increase of the helium fraction not only increases the critical mass to operate, but it also causes stronger thermal feedback due to the steeper temperature gradient in the core for the same power level and inlet conditions.
\subsubsection{Criticality \& optimization of the outlet temperature}
\begin{figure}[ht]
	\centering
	\includegraphics[width=100mm]{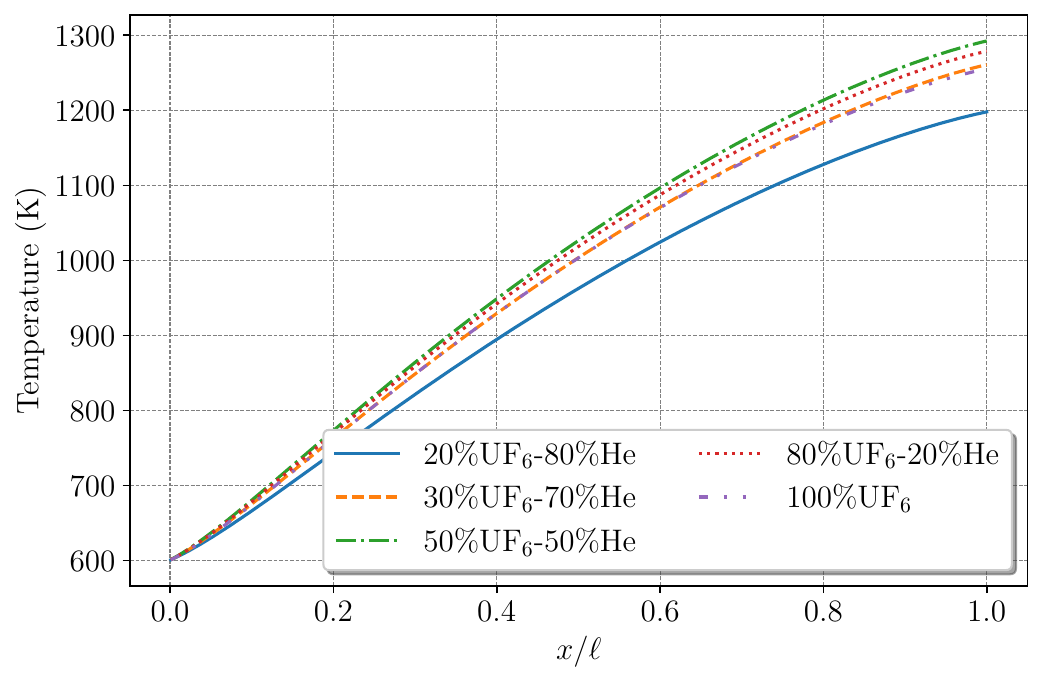}
	\caption{Evolution of the temperature in the critical subchannel for different fuel composition.}
	\label{fig:temperature_critical}
\end{figure}

The distribution of temperature in the subchannel for critical systems is shown in Fig. \ref{fig:temperature_critical}. To obtain criticality at a fixed thermal power, the inlet pressure is set as a free parameter. A Newton's algorithm iterates in order to find the pressure that yields a multiplication factor of unity.
In Fig. \ref{fig:temperature_critical} the outlet temperature of the core now varies as a function of the fuel composition. These variations can be explained using an energy balance over the system. The enthalpy of the gas mixture changes due to the thermal power, and the temperature difference $\Delta T$ between the outlet and the inlet of the system is,
\begin{equation}
	\Delta T (e) = \frac{P_{\mathrm{th}}}{A_0 u_0 \rho_0 (e) c_p (e)}.
	\label{eq:delta_t}
\end{equation}
The mass flow rate $u_0 \rho_0 A_0$ is constant thanks to the conservation of mass, Eq. \eqref{eq:flux_vector}, and $A_0$ is the cross-sectional area of the subchannel. The heat capacity $c_p$ is constant in space and is given using Mayer's relation Eq. \eqref{eq:mayer_relation}, but depends on the molar fraction of \ce{UF_6} in the mixture. The critical pressure is also a function of the molar fraction, and is fitted using an inverse power law, $p_{ref} / e^{\alpha}$, where $p_{ref}$ is the critical pressure for pure \ce{UF_6} and $\alpha$ ($=\SI{0.7628}{}$) is a fitting parameter ($r^2 = 0.99877$). The outlet temperature is then a function of the molar fraction $e$. A balance appears between density that increases with $e$ and the heat capacity $c_p$ at constant pressure per unit of mass that decreases with higher amounts of \ce{UF_6}. Replacing $\rho$ and $c_p$ with their expressions in Eq. \eqref{eq:delta_t} and maximizing the temperature difference with respect to $e$ yields,
\begin{equation}
	e_{\mathrm{opt}} = \frac{\alpha \qty(\gamma_{\ce{UF_6}} - 1)}{\qty(\alpha - 1)\qty(\gamma_{\ce{UF_6}} - \gamma_{\ce{He}})} = 0.549.
\end{equation}
A nearly equimolar mixture of fissile gas and helium is the optimal composition to obtain the highest outlet temperature at constant thermal power.
\subsection{Pressure, velocity, density and Mach number}
Pressure, velocity and Mach number distributions are calculated when a converged critical steady-state is reached in the subchannel by changing the inlet pressure. The evolution of these quantities greatly differs for different fuel compositions.
\begin{figure}[h]
	\centering
	\begin{subfigure}[b]{0.45\textwidth}
		\includegraphics[width=\textwidth]{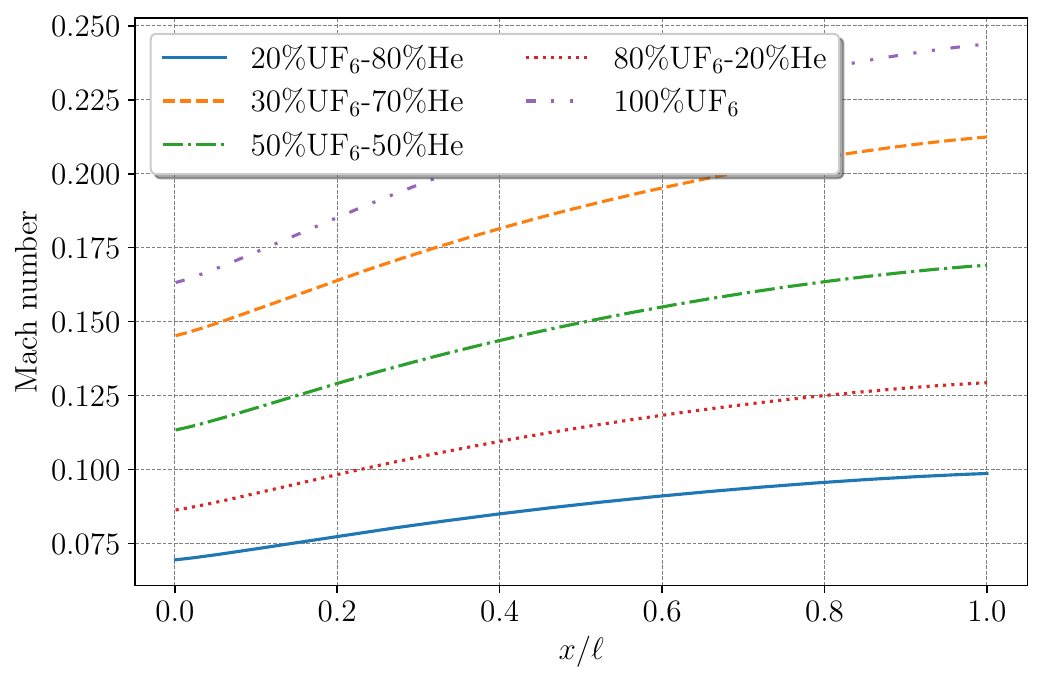}
		\caption{Mach number for different fuel mixtures along the channel.}
		\label{fig:mach}
	\end{subfigure}
	\hfill
	\begin{subfigure}[b]{0.45\textwidth}
		\includegraphics[width=\textwidth]{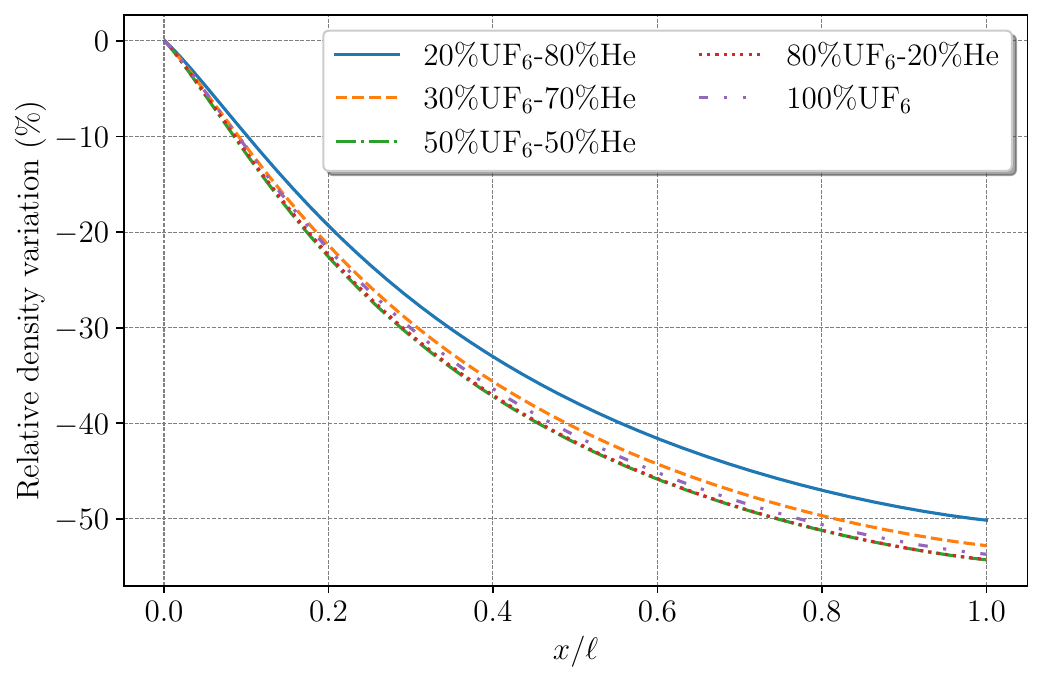}
		\caption{Relative density variations for different fuel mixtures along the channel.}
		\label{fig:density}
	\end{subfigure}
	\caption{Mach number and density distributions for a critical subchannel.}
	\label{fig:mach_and_density}
\end{figure}
The Mach number is calculated for the different fuel compositions with the speed of sound calculated using the ideal gas law,
\begin{equation}
	\mathrm{Ma} = \frac{u}{\sqrt{\bar{\gamma} \frac{p}{\rho}}},
\end{equation}
and is larger for mixtures with a lower helium content, Fig. \ref{fig:mach}. This is a consequence of the heat capacity ratio and specific gas constant being both decreasing functions of $e$. The Mach number reaches $0.25$ at the outlet of the subchannel for a mixture containing only fissile gas. Higher Mach numbers can be obtained at lower pressure or higher mass flow rates. This should be carefully considered when designing the system as the flow would experience compressibility effects above $\mathrm{Ma} = 0.3$, and because the inlet velocity is already very low for a gas flowing in a channel. Strong density variations are observed along the channel, Fig. \ref{fig:density}. The density decreases along the channel due to the increase in temperature, and the density decreases more rapidly for optimal mixtures due to higher temperature gradients, Fig. \ref{fig:temperature_mixture}.

\begin{figure}[h]
	\centering
	\begin{subfigure}[b]{0.45\textwidth}
		\includegraphics[width=\textwidth]{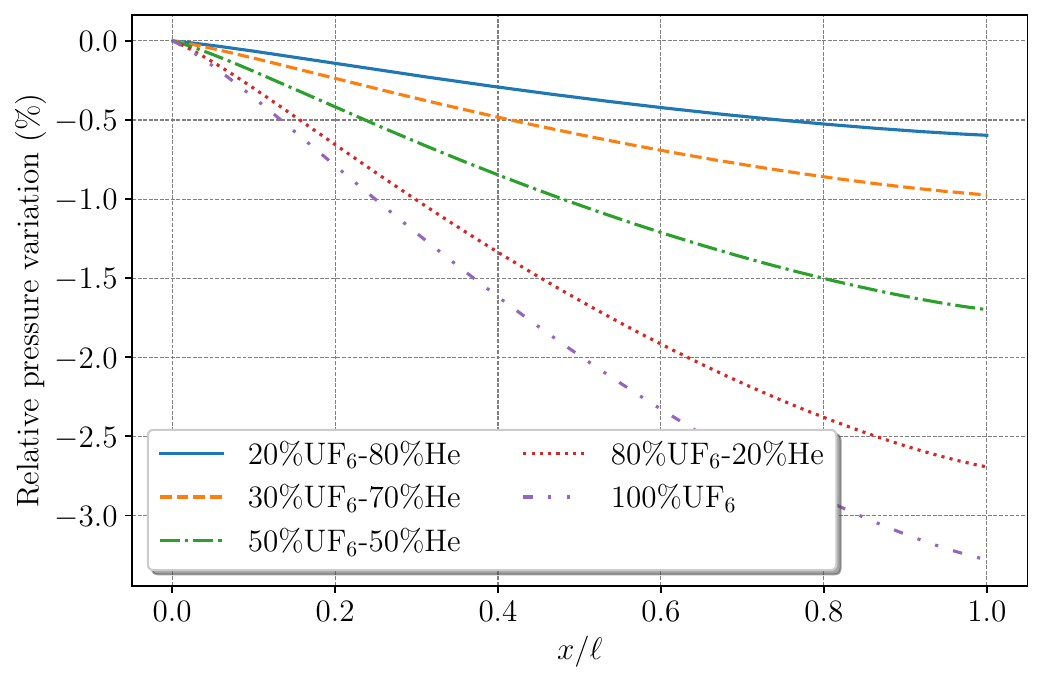}
		\caption{Relative variations of pressure for different fuel mixtures.}
		\label{fig:pressure}
	\end{subfigure}
	\hfill
	\begin{subfigure}[b]{0.45\textwidth}
		\includegraphics[width=\textwidth]{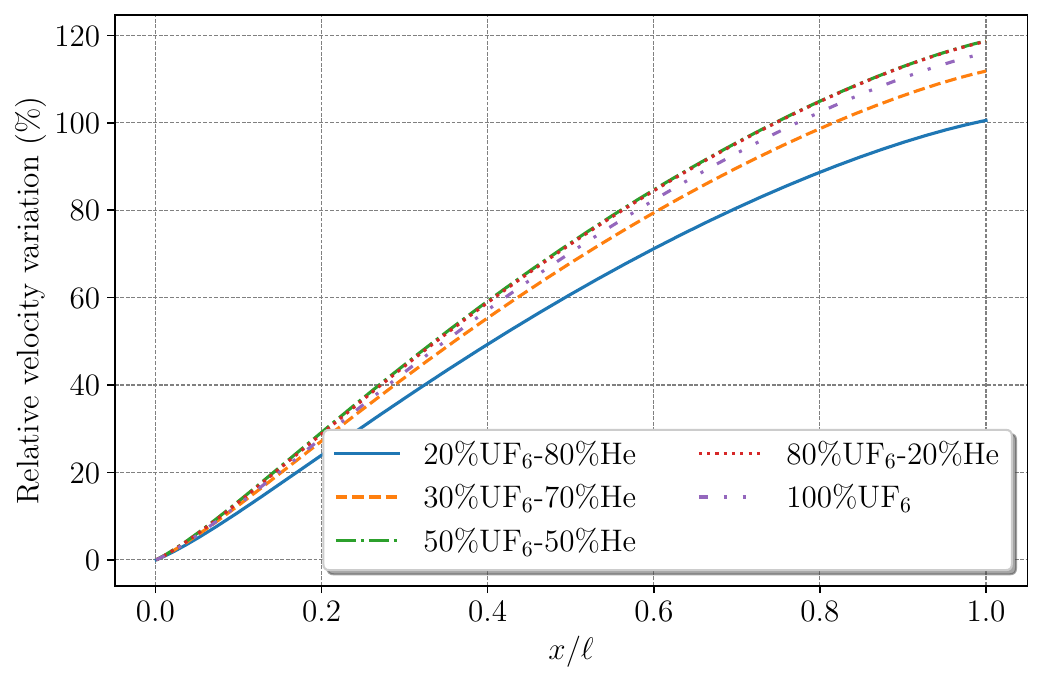}
		\caption{Relative velocity variations for different fuel mixtures.}
		\label{fig:velocity}
	\end{subfigure}
	\caption{Pressure and velocity relative distributions for different fuel mixtures.}
	\label{fig:pressure_and_velocity}
\end{figure}

On Fig. \ref{fig:pressure_and_velocity}, the relative distributions of pressure and velocity are presented. The pressure decreases along the channel, and a pressure drop of \SI{1}{\atm}, or \SI{3}{\percent} relative variation the inlet pressure is observed.
Fuels with a higher \ce{UF_6} content experience a higher pressure drop along the channel due to higher velocities attained, Eq. \eqref{eq:non_conservative_variables}.
On Fig. \ref{fig:velocity}, the fuel velocity increases along the channel and reaches up to \SI{120}{\percent} its inlet value for the equimolar mixture subjected to higher temperature gradients. This is a consequence of the mass balance equation of Eq. \eqref{eq:coupled_steady_euler}. As the temperature increases, the density decreases, and the velocity increases to maintain the same mass flow rate. The density decreases more rapidly for a higher helium content in the mixture due to higher temperature gradients, Fig. \ref{fig:temperature_mixture}.
\subsection{Feedback coefficients}
The reactivity coefficients characterizing the thermal feedback are discussed in this section. They are defined as partial derivatives of the neutron reactivity with varying thermodynamic state of the multiplying system, which is represented by the average core temperature or pressure. Derivatives are approximated by ratios of finite differences. The temperature feedback coefficient is calculated as,
\begin{equation}
	\alpha = \frac{\varrho_{\mathrm{pert}} - \varrho_{\mathrm{nom}}}{\langle T_{\mathrm{pert}} \rangle - \langle T_{\mathrm{nom}} \rangle},\quad T_{\mathrm{pert}} = T_{\mathrm{nom}} \qty(1 + \epsilon),
	\label{eq:feedback_coefs}
\end{equation}
where $\varrho$ denotes the static reactivity defined as $1 - 1 / \keff$. The brackets represent the mean of the considered distribution over the core, $\epsilon$ is a perturbation parameter taken to be equal to $\num{1e-6}$.
\begin{table}[ht]
	\centering
	\caption{Feedback coefficients for different mixtures.}
	\vspace{5mm}
	\begin{tabular}{cc}
		\hline
		Mixture                       & $\alpha$ (\SI{}{\pcm\per\kelvin}) \\
		\hline
		\hline
		$100\%$\ce{UF_6}              & $\num{-20.2}$                     \\
		$80\%$\ce{UF_6}-$20\%$\ce{He} & $\num{-25.2}$                     \\
		$50\%$\ce{UF_6}-$50\%$\ce{He} & $\num{-46.1}$                     \\
		$30\%$\ce{UF_6}-$70\%$\ce{He} & $\num{-73.5}$                     \\
		$20\%$\ce{UF_6}-$80\%$\ce{He} & $\num{-92.4}$                     \\
		\hline
	\end{tabular}
	\quad
	\begin{tabular}{cc}
		\hline
		Mixture                       & $\delta$ (\SI{}{\pcm\per\atm}) \\
		\hline
		\hline
		$100\%$\ce{UF_6}              & $\num{437}$                    \\
		$80\%$\ce{UF_6}-$20\%$\ce{He} & $\num{618}$                    \\
		$50\%$\ce{UF_6}-$50\%$\ce{He} & $\num{1084}$                   \\
		$30\%$\ce{UF_6}-$70\%$\ce{He} & $\num{2062}$                   \\
		$20\%$\ce{UF_6}-$80\%$\ce{He} & $\num{3570}$                   \\
		\hline
	\end{tabular}
	\label{table:feedbacks}
\end{table}
Fig. \ref{table:feedbacks} shows that a growth in the helium proportion increases the magnitude of the thermal feedback coefficients. This can be explained using Eq. \eqref{eq:cv_mixture} and Fig. \ref{fig:temperature_mixture}. As the proportion in helium increases, the heat capacity decreases, resulting in higher temperature gradients for the same thermal power. As the fuel density is proportional $1/T$, the larger the temperature gradient the lower the density. This leads to an overall reduction of the macroscopic cross-sections as shown by Eq. \eqref{eq:sigma_p_t}. In this analysis, the pressure contribution is neglected as the relative pressure variations along the channel are less than $\SI{2.1}{\percent}$. A decrease of the macroscopic cross-sections ultimately induces a loss of reactivity, much bigger at higher temperatures. Replacing the temperature by the pressure distribution in Eq. \eqref{eq:feedback_coefs} allows computing pressure coefficients, named $\delta$ as shown in Fig. \ref{table:feedbacks}.
Fig. \ref{table:feedbacks} confirms the trend observed in Fig. \ref{fig:keff_vs_p}. For a reference point of \SI{40}{\atm}, changes in reactivity are larger for a fuel gas with a higher helium proportion. An increase of the pressure is a way to insert reactivity in such system.

\section{Discussion \& Conclusion}
This work explored the physical behavior of a stationary ideal gas fuel reactor. The fuel mixture was composed of enriched uranium hexafluoride and helium flowing in a graphite subchannel coated with beryllium oxide. The addition of helium, an inert gas, served as a practical means to reduce the heat capacity of the fuel, allowing for a higher outlet temperature at constant thermal power without reacting with the core structure or activating under irradiation.

A critical subchannel can be reached by varying the inlet pressure of the reactor while keeping the thermal power constant. An increase in pressure brings the mass of the fuel closer to the critical mass by increasing its density. The use of helium in the mixture increases the critical pressure due to dilution of fissile materials and higher temperature gradients. The system exhibits high temperature feedback coefficients due to a fuel thermal expansion coefficient equal to the inverse of temperature. A pressure growth induces an increase in reactivity.

However, non-ideality of the gas mixture should be taken into account, as well as the chemistry of fuel. Moreover, the evolution of the gas fuel was not studied, and above \SI{1500}{\kelvin} or under irradiation \ce{UF_6} starts disassociating, changing the fuel composition \cite{clement1976analysis}. Although the disassociation of \ce{UF_6} is an important reaction under irradiation, it has been shown experimentally that the reverse reaction, i.e. recombination with fluorine, allows an equilibrium concentration of \ce{UF_6} to exists even at low temperatures \cite{DMITRIEVSKII1960185}. Corrosion is also an important topic, as \ce{UF_6} is a very aggressive chemical, especially at high temperatures. A study conducted on ceramics exposed to high temperature \ce{UF_6} concluded that \SI{1273}{\kelvin} is the maximum compatible temperature for a \ce{Al_2 O_3} refractory ceramic \cite{ceramic_corrsion}. Therefore, the lifetime of the channel should be investigated in more details. Non-conventional heat extraction mechanism such as magneto-inductive of magneto-hydrodynamic conversion could be interesting topic for electricity production. As future work, the neutron transport solver will be extended to multiple energy groups, allowing for a better estimation of scalar fluxes. Possible pressure losses due to friction in the subchannel should also be investigated.
\bibliography{sn-bibliography.bib}

\end{document}